\documentclass[pre,twocolumn,showpacs,amsmath,amssymb,floatfix]{revtex4}

\usepackage{graphicx}
\usepackage{dcolumn} 
\usepackage{bm}      
\usepackage{color}
\begin{document}

\title{Finite dissipation and intermittency in magnetohydrodynamics}
\author{P.D. Mininni$^{1,2}$ and A. Pouquet$^1$}
\affiliation{$^1$ Departamento de F\'\i sica, Facultad de Ciencias Exactas y
         Naturales, Universidad de Buenos Aires, Ciudad Universitaria, 1428
         Buenos Aires, Argentina. \\
             $^2$ NCAR, P.O. Box 3000, Boulder, Colorado 80307-3000, U.S.A.}
\date{\today}

\begin{abstract}
We present an analysis of data stemming from numerical simulations of decaying magnetohydrodynamic (MHD) turbulence up to grid resolution of $1536^3$ points and up to Taylor Reynolds number of $\sim 1200$. The initial conditions are such that the initial velocity and magnetic fields are helical and in equipartition, while their correlation is negligible. Analyzing the data at the peak of dissipation, we show that the dissipation in MHD seems to asymptote to a constant as the Reynolds number increases, thereby strengthening the possibility of fast reconnection events in the solar environment for very large Reynolds numbers. Furthermore, intermittency of MHD flows, as determined by the spectrum of anomalous exponents of structure functions of the velocity and the magnetic field, is stronger than for fluids, confirming earlier results; however, we also find that there is a measurable difference between the exponents of the velocity and those of the magnetic field, as observed recently in the solar wind. Finally, we discuss the spectral scaling laws that arise in this flow.
\end{abstract}
\maketitle

As observations of astrophysical flows become more detailed, both spatially and temporally, the need for a deeper understanding of turbulent flows grows. In many such flows the fluid is coupled to a magnetic field the dynamics of which can be understood in the magnetohydrodynamic (MHD) approximation granted the analysis is confined to the large scales and under the hypothesis that the velocities are substantially smaller than the speed of light so that the displacement current in Maxwell's equations can be neglected. This latter condition is easily fulfilled, characteristic bulk velocities in the solar wind being typically between 400 and 800 km s$^{-1}$, and turbulent velocities in the solar convection zone being $\sim 1$ km s$^{-1}$. The MHD approach breaks down at small scales where kinetic plasma effects become important and one needs to include other terms in a generalized Ohm's law, such as ambipolar diffusion in weakly ionized plasmas as encountered in the interstellar medium, the Hall current for highly ionized media such as the solar wind, or an anisotropic pressure tensor. In such cases, the nonlinearities of the dynamical equations become more numerous and complex, parameter space is expanded and the resulting problem is quite challenging. For that reason, MHD is still a valid approach, albeit a simplified one, to tackle questions concerning the fate of a turbulent fluid coupled to magnetic fields.

Laboratory experiments have classically been one venue to understand the physics of such fluids, for example in the context of reconnection \cite{gekelman}. Using liquid metals in the laboratory is a challenge for exploring the high magnetic Reynolds number $R_M$, the governing parameter of the problem, because the magnetic Prandtl number $P_M=\nu/\eta$ is small, typically $10^{-6}$ for sodium ($\nu$ and $\eta$ are the viscosity and magnetic resistivity). A dynamo has been obtained recently within a turbulent flow \cite{bourgoin} but the high $R_M$ regime, as is the case for astrophysical flows, remain unattainable in the laboratory. On the other hand, {\it in situ} observations of the Earth environment have grown in importance recently, e.g. with the multi-spacecraft mission CLUSTER \cite{cluster,recent_nature}. Observations are quite complex but indicate clearly several features, such as power law energy spectra \cite{goldstein80} and intermittency \cite{burlaga} (see e.g. \cite{tu_marsch} for review). One of the issues is to assess what kind of scaling laws obtains for both the velocity and the magnetic field; moreover, the flow may develop an anisotropic weak turbulence spectrum at small scale as shown recently in direct numerical simulations (DNS) \cite{second} and as observed in the magnetosphere of Jupiter \cite{saur}.

Indeed, DNS may help but remain challenging in three space dimensions (3D). A plethora of results concerning energy spectra in MHD have emerged recently, with different power laws in different regions of parameter space, although the boundaries between these regions are not fully understood and more exploration remains to be done. However, whatever the inertial index of the spectrum, one may ask whether, for correlation functions of higher order, similarities between hydrodynamic and MHD turbulence persist. It is already known that it does not in two space dimensions (2D) \cite{politano,biskamp,carbone}, MHD being more intermittent than neutral fluids but the data in 3D remains scarce.

A further problem concerns the dissipation of energy in the limit of high Reynolds number $R_e$. Mathematically, this is an open problem in 3D for fluids and MHD, and yet it is central for astrophysics where dissipative structures, reconnection and acceleration of particles are well observed \cite{recent_nature}. Intermittency (as measured by anomalous exponents of structure functions) and singular behavior are linked since the latter (except for a thin boundary layer delimiting the thickness of the structure) is likely to occur on a set of strong small-scale fluid elements highly localized spatially, be it vortex filaments or current and vorticity sheets. We thus propose in this paper an assessment of dissipation, small scale structures, intermittency, and scaling laws, by analyzing a flow computed up to a grid resolution of $1536^3$ points.

The incompressible MHD equations read:
\begin{eqnarray}
&& \partial_t {\bf v} + {\bf v} \cdot \nabla {\bf v} = 
    -\rho_0^{-1} \, \nabla {\cal P} + {\bf j} \times {\bf b} + 
    \nu \nabla^2 {\bf v} , 
\label{eq:MHDv} \\
&& \partial_t {\bf b} = \nabla \times ( {\bf v} \times
    {\bf b}) +\eta \nabla^2 {\bf b} \ ,
\label{eq:MHDb}
\end{eqnarray}
with ${\bf v}$ the velocity, ${\bf b}$ the magnetic field, ${\cal P}$ the pressure, $\rho_0=1$ the (uniform) density, ${\bf j}=\nabla \times {\bf b}$ the current density, and ${\bf \nabla} \cdot {\bf v} = \nabla \cdot {\bf b} = 0$. When $\nu=\eta=0$, the energy $E=\left<v^2+b^2\right>/2$, magnetic helicity $H_b=\left<{\bf A} \cdot {\bf b}\right>$ (with ${\bf A}$ the vector potential such that ${\bf b} = \nabla \times {\bf A}$), and cross helicity $H_C=\left<{\bf v} \cdot {\bf b}\right>/2$, are conserved. We solve Eqs. (\ref{eq:MHDv}) and (\ref{eq:MHDb}) in a 3D box using periodic boundary conditions and a pseudospectral method dealiased by the standard 2/3 rule; $k_\textrm{min}=1$ for a box of length $L_0=2\pi$, and $N$ regularly spaced grid points lead to a maximum wavenumber $k_\textrm{max}=N/3$. At all times, we preserve $k_D/k_\textrm{max}<1$, where $k_D$ is the dissipation wavenumber. 

The initial conditions are constructed from a superposition of Beltrami flows from wavenumbers $k=1$ to 3, to which smaller-scale random fluctuations with a spectrum $\sim k^{-3}\exp [-2(k/k_0)]^2$ for $k>3$ are added (see \cite{Mininni06}). The phases of the modes with $k>3$ are chosen from a Gaussian random number generator in such a way that the initial cross-correlation of the two fields is negligible: initially, $E_V = E_M = 0.5$, 
$H_C\sim 10^{-4}$, 
and $H_M  \sim 0.45$. Resolutions of runs described in this paper range from $N=64$ to $N=1536$ (see Table \ref{table:runs}). The largest resolution run is stopped close to the peak of dissipation, $t=3.7$; its initial quasi-ideal phase is described in \cite{Mininni06} and the total energy spectra that develop, together with the ensuing anisotropy of the small scales, is given in \cite{second}. Near the peak of dissipation, the Reynolds number based on the integral scale of the flow velocity is $R_e = UL/\nu \approx 9200$, and that based on the Taylor scale is $R_\lambda = U\lambda /\nu \approx 1700$; $U$ is the r.m.s. velocity, the integral scale is defined as $L = 2\pi E^{-1} \int k^{-1} E(k)dk $ and the Taylor scale as $\lambda = 2\pi ( E / \int k^2 E(k)dk )^{1/2} $, with $E(k)$ the total energy spectra.

\begin{table}
\caption{
Runs, linear resolution $N$, viscosity $\nu$ and magnetic diffusivity $\eta$, Reynolds number $Re$, and Taylor Reynolds number $R_\lambda$ at peak of dissipation.}
\begin{ruledtabular}
\begin{tabular}{ccccc}
Run & $N$  & $\nu=\eta$            & $Re$  & $R_\lambda$   \\
\hline
I   & 64   & $8 \times 10^{-3}$    & 390   & 180    \\
II  & 128  & $3 \times 10^{-3}$    & 790   & 280   \\
III & 256  & $1.25 \times 10^{-3}$ & 1600  & 430  \\
IV  & 512  & $6 \times 10^{-4}$    & 3100  & 630    \\
V   & 1536 & $2 \times 10^{-4}$    & 10500 & 1180
\label{table:runs} \end{tabular} \end{ruledtabular} \end{table}

\begin{figure}
\includegraphics[width=8.5cm]{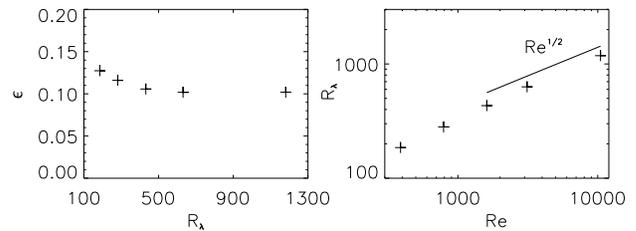}
\caption{{\it Left:} Total energy dissipation rate $\epsilon$ as a function of Taylor Reynolds number $R_\lambda$ for several runs with the same initial conditions. For large $R_\lambda$, $\epsilon$ seems to be independent of $R_\lambda$. {\it Right:} $R_\lambda$ as a function of $R_e$ calculated at the peak of dissipation for the same runs. 
The straight line indicates the classical turbulent scaling $R_\lambda \sim R_e^{1/2}$.}
\label{fig:dissipation} \end{figure}

We focus on the fully developed turbulent regime close to the peak of dissipation. Figure \ref{fig:dissipation} gives the variation of the maximum of the total energy dissipation rate $\epsilon = \nu\left<\omega^2\right> + \eta \left<j^2\right>$  with $R_\lambda$ ($\mbox{\boldmath $\omega$}=\nabla \times {\bf v}$ is the vorticity) for the runs of Table \ref{table:runs}. For large $R_\lambda$, $\epsilon$ seems to become independent of $R_\lambda$. This result is not entirely unexpected. On the one hand, the dissipation of energy is known to tend to a constant  in the case of neutral fluids (${\bf b}\equiv 0$) \cite{kaneda}; and when restricting the MHD dynamics to 2D (which, to lowest order, is the evolution that is expected in the presence of a  strong uniform magnetic field), the energy dissipation was shown similarly to be constant \cite{politano, biskamp}; in 3D MHD, an indication that this may be the case as well was obtained for the Orszag-Tang vortex \cite{Mininni06} although at a lower resolution ($512^3$ points) and Reynolds number ($R_e\sim 5600$) which did not allow for a clear scaling. Here, it appears that we have reached the beginning of an asymptotic regime where dissipation is constant and the Taylor Reynolds number scales as the square root of the Reynolds number $R_e$ (see Fig. \ref{fig:dissipation}), as expected for a fully developed turbulent flow.

\begin{figure}
\includegraphics[width=8.4cm]{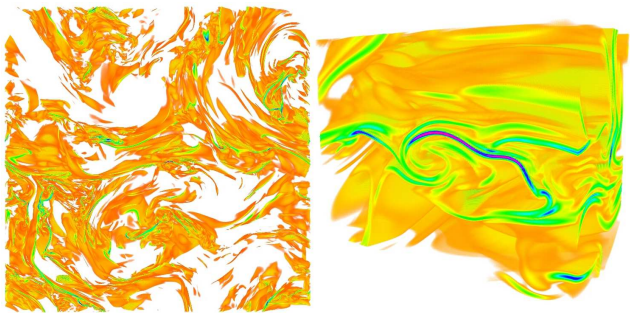}
\caption{(Color online) Current density in a slice of the full box (left), and in a subregion (right) showing folding and rolling of the current sheets. Vorticity organizes in the same fashion, although current sheets are thinner. The high intensity (dark, green and magenta) is concentrated in a thin layer within subvolumes of the flow.}
\label{fig:structures} \end{figure}

For $\epsilon$ to remain constant with vanishing viscosity and resistivity, one can think of several scenarios; either we have intense dissipative structures that are more space-filling as $R_e$ grows, or else the structures remain sparse but become very sharp. Both may be happening, with a myriad of current sheets of intermediate to large intensity, and a few very sharp structures. When plotting the histogram of one component of the current intensity (not shown), one observes that, as the Reynolds number increases, the wings of the PDF stabilize at intermediate values but substantially higher extrema are reached. Fig. \ref{fig:structures} gives a 3D rendering of the current density in a slice of the entire domain, and in a subvolume showing folding and rolling of the current sheet. Visualizations of the time evolution of these structures confirm that the rolling takes place as the result of a Kelvin-Helmholtz-like instability as observed in the solar wind \cite{recent_nature}.

\begin{figure}
\includegraphics[width=8.5cm]{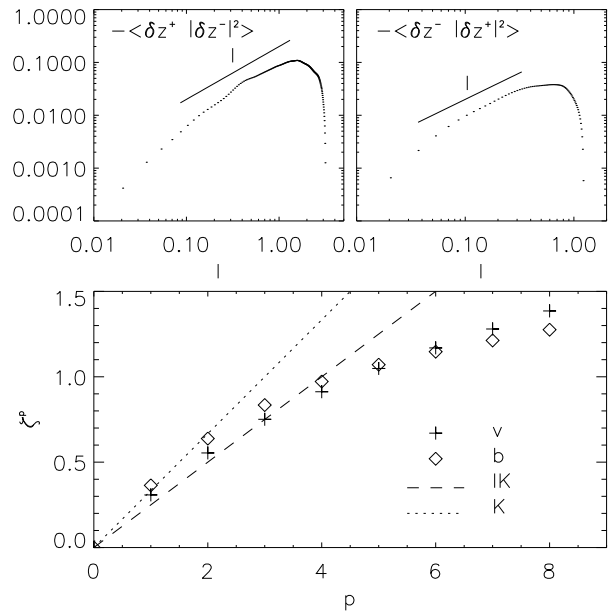}
\caption{{\it Top:} $-\left< \delta z_\parallel^\pm (l)|\delta {\bf z}^\mp (l)|^2 \right>$ as a function of the displacement $l$ in the fully developed turbulent regime of the $1536^3$ run. A slope of $1$ is indicated as a reference. {\it Bottom:} Scaling exponents $\zeta^p$ for the velocity and magnetic fields. The Iroshnikov-Kraichnan (IK) and Kolmogorov predictions are shown as a reference. Although $\zeta^2_v$ is closer to Kolmogorov scaling, $\zeta^3_v$ is far from the hydrodynamic value of 1. Note the measurable difference between both sets of exponents.}
\label{fig:theorem}
\end{figure}

One way to determine the statistics of such structures is to examine the behavior of structure functions; at order p for a field ${\bf u}$, they are defined as $\left<|\delta u_\parallel(l)|^p\right>$, with $\delta u_\parallel(l)=u_\parallel({\bf x+l})-u_\parallel({\bf x})$ with homogeneity and isotropy assumed and with $u_{\parallel}$ the longitudinal component of the vector ${\bf u}$ that projects along ${\bf l}$. Assuming self similarity leads to $\left<|\delta u_\parallel(l)|^p\right>\sim l^{\zeta_u^p}$, with $\zeta_u^p=ap$ for a scale invariant (non-intermittent) field ($a=1/3$ for Kolmogorov scaling, $a=1/4$ for Iroshnikov-Kraichnan scaling). Departures from such a linear scaling are observed experimentally, observationally and numerically but a normal (linear) scaling occurs for third-order functions, expressing the conservation laws of the ideal case: total energy and cross-correlation \cite{Politano98}, as well as magnetic helicity \cite{politano_HM}. In terms of the Els\"asser variables ${\bf z}^{\pm}={\bf v}\pm {\bf b}$, the first two conservation laws lead to
\begin{equation}
\left< \delta z_\parallel^\mp (l) \left| \delta {\bf z}^\pm (l)
    \right|^2 \right> = -\frac{4}{3} \epsilon^\pm l ,
\label{eq:theorem}
\end{equation}
where $\epsilon^\pm$ are the dissipation rates of ${\bf z}^\pm$. From these expressions, the flux of total energy $\epsilon$ and of the cross correlation between the fields $\epsilon_C$ can be computed as a function of the scale $l$. The relations given by Eq. (\ref{eq:theorem}) as evaluated directly from the $1536^3$ data near the peak of dissipation are shown in Fig. \ref{fig:theorem}. A linear dependence with $l$ is observed in a range of scales for both flux functions although the scaling is slightly better for the $\epsilon^-$ flux; as a result, this is the quantity we will use for the extended self-similarity (ESS) analysis \cite{Benzi93b}: in fluid turbulence, it is a common practice to plot structure functions in terms of each other, the third order one being particularly relevant since it is proportional to $l$ and can be used to define the inertial range and to improve the estimation of the scaling exponents. We thus determined the anomalous scaling exponents for MHD for the Els\"asser variables \cite{second} using ESS. We show here the determination of these exponents for the velocity and magnetic field, assuming isotropy as before. A measurable difference is obtained, as observed recently in the solar wind \cite{Podesta}; it corresponds to a steeper magnetic energy spectrum (close to Kolmogorov scaling) and a shallower kinetic energy spectrum (close to Iroshnikov-Kraichnan scaling). Indeed, for the second order scaling exponent of the velocity field $\zeta_v^2 = 0.55\pm0.01$, and for the magnetic field $\zeta_b^2 =0.64\pm0.01$. These exponents in turn lead to a kinetic energy spectrum $E_v(k) \sim k^{-1.55}$ and a magnetic energy spectrum $E_M(k) \sim k^{-1.64}$. Note that for both fields, $\zeta^3\not=1$ indicating already at third order a departure from Kolmogorov phenomenology, and that the intermittency is stronger. However, in this simulation the exponents of the Els\"asser variables ${\bf z}^\pm$ are closer to Iroshnikov-Kraichnan scaling than to Kolmogorov (with the second order exponent near $0.6$ \cite{second} because of intermittency corrections). Note that the different exponents that have been observed in the solar wind \cite{Podesta} are on the average $E(k) \sim k^{-1.6}$, $E_v(k) \sim k^{-1.5}$, and $E_b(k) \sim k^{-1.66}$ \cite{Podesta}; in solar active regions, variations have also been measured with a monotone decrease of the exponent at a given order when the strength of the flare augments (from M1 to X1) \cite{abramenko07}.

\begin{figure}
\includegraphics[width=8.5cm]{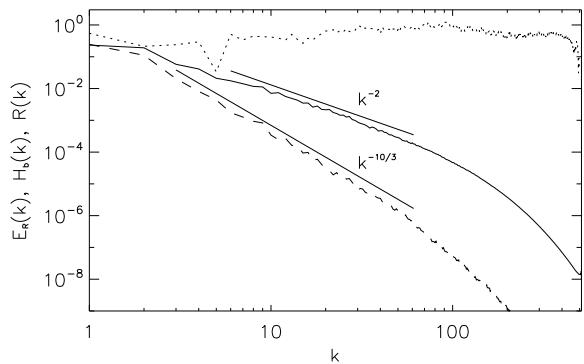}
\caption{Residual energy $E_R(k)$, 
magnetic helicity $H_b(k)$ 
and ratio $R(k)$ (see text) at the peak of dissipation in run V.}
\label{fig:scaling}
\end{figure}

The different scaling of the velocity and the magnetic field can thus be explained in terms of the different intermittency properties of each field. Indeed, in MHD turbulence current sheets are thinner than vortex structures, a property that results in faster dissipation of magnetic energy than of mechanical energy. The development of thin structures in the current in turn leads to a steeper spectrum for the magnetic field than for the velocity field. Other scaling laws arise in the flow, specially at very high Reynolds number (run V), that in some cases have been previously reported in observations or predicted using theoretical arguments. Figure \ref{fig:scaling} shows the residual energy spectrum $E_R(k)=E_b(k)-E_v(k)$ \cite{grappin}, with for a Iroshnikov-Kraichnan scaling for  total energy should scale as $k^{-2}$ and for Kolmogorov scaling goes as $k^{-7/3}$. The residual energy spectrum in the simulation is consistent with $k^{-2}$. The magnetic helicity $H_b$ seems to follow a $k^{-10/3}$ spectrum. This scaling has also been observed in the inverse cascade range of the magnetic helicity \cite{Mueller} and is not well understood. It could result from the Alfv\'enic balance between $E_b(k)/E_v(k)$ and $k^2 H_b/H_v$ \cite{Mueller} where $H_v = \left< {\bf v}\cdot \mbox{\boldmath $\omega$}\right>$ is the kinetic helicity and where the factor $k^2$ follows from dimensional reasons; Fig. \ref{fig:scaling} also shows $R(k) = [E_b(k)/E_v(k)] [k^2 H_b(k)/H_v(k)]^{-1}$ and although such a balance is plausible, a slow increase of $R(k)$ with $k$ can also be observed.

The measurement of the energy input $\epsilon$ (and ensuing heating) in magnetospheric plasmas is an outstanding problem that the present CLUSTER mission helps unravel and that future missions, such as the Magnetospheric MultiScale (MMS) to be launched in 2013 is designed to study. From solar wind observations, it can be measured by using the exact scaling laws used here, which seem to be reasonably followed by the numerical data. The fact that the present study shows the constancy of $\epsilon$ with Reynolds number indicates that energy is transferred efficiently to small scales in MHD as long as sufficient scale separation is available. Kinetic plasmas effects will come into play as the cascade meets, e.g., the ion-cyclotron frequency, leaving open the issue of what follows at smaller scale, but the present results imply that energy can be cascaded rapidly (independently of the Reynolds number) to the smallest available scales by MHD turbulence.
Moreover, dissipation is achieved in localized regions with strong magnetic field gradients, in the form of current sheets. These extreme events, more probable at small scales than what is expected from a normal distribution, represent a break down of scale invariance and give rise to intermittency. The thin current sheets result in a more intermittent magnetic field than velocity field, and in turn make the magnetic energy spectrum steeper than the kinetic energy spectrum. Remarkably, the second order scaling exponents and spectral indices for the kinetic, magnetic, and total energy in the MHD simulation at largest Reynolds number are in good agreement with the ones reported for the solar wind. However, care must be taken when extracting conclusions about scaling laws in MHD turbulence. Simulations have been reported where the total energy spectrum follows different power laws depending on properties of the forcing \cite{Matthaeus}, and in the solar wind variations in the total energy spectrum from $\sim k^{-3/2}$ to $\sim k^{-5/3}$ have been observed \cite{Podesta}. It is not our intention to say that MHD turbulence has unique scaling properties represented by our simulations, but rather that the determination of scaling laws in MHD turbulence, and the explanation of the results from solar wind observations, require the study of often neglected phenomena as intermittency, and the measurement of high order statistics of the velocity and the magnetic fields.

{\it The largest run was performed through a BTS grant at NCAR, sponsored by NSF. PDM acknowledges support from grant UBACYT X468/08 and from the Carrera del Investigador Cient\'{\i}fico of CONICET. Three-dimensional visualizations use VAPOR, a software for interactive visualization and analysis of terascale datasets \cite{Clyne07}.}

\end{document}